\documentclass[10pt,a4paper]{article}
\usepackage[T2A]{fontenc}
\usepackage[utf8]{inputenc}
\usepackage[english]{babel}
\usepackage{amssymb,graphicx}
\usepackage{amsmath,amsfonts}
\usepackage{amsthm}
\usepackage{fullpage}
\usepackage{hyperref}
\title{Generalized Jack polynomials and the AGT relations for the
  $SU(3)$ group}
\author{S.~Mironov$^{a,b}$\thanks{sa.mironov\_1@physics.msu.ru} ,
  And.~Morozov$^{a,c}$\thanks{andrey.morozov@itep.ru} ,
  Y.~Zenkevich$^{a,b}$\thanks{yegor.zenkevich@gmail.com,
    zenkevich@ms2.inr.ac.ru},\\
  {\small $^a$\textit{ITEP, Moscow, Russia}}\\
  {\small $^b$\textit{Institute for Nuclear Research of the Russian
      Academy of
      Sciences, Moscow, Russia}}\\
  {\small $^c$\textit{Moscow State University, Physics Department,
      Moscow, Russia}}} \date{}
\begin{document}
\maketitle
\vspace{-55ex}
\begin{flushright}
  ITEP-TH-49/13
\end{flushright}
\vspace{42ex}

\begin{abstract}
  We find generalized Jack polynomials for the $SU(3)$ group and
  verify that their Selberg averages for several first levels are
  given by Nekrasov functions. To compute the averages we derive
  recurrence relations for the $\mathfrak{sl}_3$ Selberg integrals.
\end{abstract}

\section{Introduction}
\label{sec:introduction}
The AGT relations~\cite{AGT} provide an extremely interesting link
between the four dimensional $\mathcal{N}=2$ gauge theories of class
$S$~\cite{Gaiotto} and two dimensional conformal field
theories. Moreover, these relations offer a new view on a variety of
related fields, such as integrable systems~\cite{IntSyst}, matrix
models~\cite{Matrix},~\cite{AGTproof}, etc.

The most surprising property of the AGT relations is that they give an
additional and unexpected structure on the conformal field theory
Hilbert space. One usually writes the states in this space as
descendants of some primary field: $L_{-Y}| \alpha \rangle$, all the
correlators therefore become sums over Young diagrams $Y$. However,
the Nekrasov function is a sum over \emph{pairs} of Young diagrams
$\vec{Y}$. Finding the corresponding basis $| \vec{Y} \rangle$ in CFT
is an interesting problem. This basis was found explicitly in the case
of $c=1$~\cite{BB} and it was argued to exist in the general
case~\cite{AFLT}. Concretely, if one performs the bosonization of the
Virasoro algebra, the basis vectors are expressed through the
generalized Jack polynomials which are defined as the polynomial
eigenfunctions of a certain differential operator\footnote{It is
  suspected that generalized Jack polynomials are actually common
  eigenfunctions of \emph{an infinite family} of differential
  operators.}~\cite{MS}.

One can also compute the correlators in CFT using the Dotsenko-Fateev
approach, in which they are given by certain multiple integrals
related to the Selberg integrals~\cite{Selberg}. In this setting the
generalized Jack polynomials also play a distinguished role: their
Selberg averages are factorised into a product of linear functions of
momenta. More precisely, the averages are given by the Nekrasov
functions.

More generally, the AGT relations map the gauge theory with the
$SU(N)$ group to the four point conformal block in the Toda field
theory with two general and two degenerate fields~\cite{SU(3)}. The
same factorisation of the $\mathfrak{sl}_N$ Selberg averages should
happen in this case as well.

In this Letter we explicitly find generalized Jack polynomials for the
group $SU(3)$ and check that their Selberg averages indeed reproduce
the Nekrasov functions on the first levels. To compute the averages we
derive the $W$-constraints for the $\beta$-deformed $A_2$ quiver
matrix model. In section~\ref{sec:diff-oper} we introduce the
differential operator whose polynomial eigenfunctions are given by the
generalized Jack polynomials, compute them explicitly and check their
elementary properties. In section~\ref{sec:norm-compl} we derive the
Dotsenko-Fateev representation of the conformal block in Toda field
theory and show that the AGT relations hold if certain Selberg
averages of generalised Jack polynomials are given by the Nekrasov
functions. Using the $W$ constraints we compute the averages and check
the relevant formulas for the first levels. The Nekrasov functions and
AGT relations are provided in
Appendix~\ref{sec:nekrasov-function}. The $W$ constraints are
presented in Appendix~\ref{sec:w-viras-constr}.

\section{Differential operator}
\label{sec:diff-oper}
Generalized Jack polynomials $J_{\vec{\lambda}} (\{ p_k^{(a)} \}|\beta,
\{ a_a \})$ are labelled by an $N$-tuple of Young
diagrams $\vec{\lambda} = \{ \lambda_{(1)}, \lambda_{(2)}, \ldots\}$. They are eigenfunctions of the differential
operator $\hat{D} = \sum_{a=1}^N \hat{H}_a + \sum_{a < b}
\hat{H}_{ab},$ where
\begin{gather}
  2\hat{H}_a =  \sum_{n,m \geq 0} \left( \beta (n + m) p_n^{(a)}
    p_m^{(a)} \frac{\partial}{\partial p_{n+m}^{(a)}} + n m p_{n+m}^{(a)}
    \frac{\partial^2}{\partial p_n^{(a)} \partial p_m^{(a)}} \right) + \sum_{n
    \geq 1} \left( 2a_a + (1- \beta) (n-1) \right) n p_n^{(a)}
  \frac{\partial}{\partial p_n^{(a)}}  , \notag\\
  \hat{H}_{ab} = (1-\beta) \sum_{n \geq 1} n^2 p_n^{(b)}
  \frac{\partial}{\partial p_n^{(a)}}\,. \notag
\end{gather}
with eigenvalues $\varkappa_{\vec{\lambda}} = \sum_{a=1}^N
\sum_{(i,j) \in \lambda_a} (a_a - \beta(i-1) + (j-1))$.

This definition allows one to find first few polynomials as linear
combinations of $p_k^{(a)}$. Since the eigenvalues
$\varkappa_{\vec{\lambda}}$ are known, the problem of finding the
eigenfunctions reduces to a system of linear algebraic equations. The
elementary properties of the generalized Jack polynomials are:
\begin{enumerate}
\item \textbf{Orthogonality and normalization.} If one sets
  $\widetilde{J}_{\vec{A}} (p) = \prod_{a=1}^{N} m_{A_a}(p)+ \ldots$
  with $m_A$ being the monomial symmetric function then the family of
  generalized Jack polynomials is orthonormal, $\langle
  \widetilde{J}^{*}_{\vec{A}} , \widetilde{J}_{\vec{B}} \rangle =
  \delta_{\vec{A}\vec{B}}$, with respect to the scalar product
\begin{equation}
  \langle f(p_k) , g (p_k) \rangle = \left. f \left( \frac{n}{\beta}
  \frac{\partial}{\partial p_k} \right) g(p_k) \right|_{p_k = 0} \notag
\end{equation}
Here the conjugate polynomials are defined as follows:
\begin{equation}
  \label{eq:16}
  \widetilde{J}^{*}_{A_{1},A_{2},\ldots, A_{N}}(p^{(1)},\ldots,p^{(N)}|\beta, a_1,
  \ldots ,a_N) = \widetilde{J}^{*}_{A_{N},A_{N-1},\ldots, A_{1}}(p^{(N)},\ldots,p^{(1)}|\beta, a_N,
  \ldots ,a_1)\,.
\end{equation}
It will be more convenient for us to use a different normalization
\begin{equation}
  \label{eq:23}
  J_{\vec{A}}(p) = (-1)^{|\vec{A}|} \beta^{-2|\vec{A}|}
  \prod_{a<b} g_{A_b , A_a}(a_b - a_a) \widetilde{J}_{\vec{A}}(p)\,,
\end{equation}
where $g_{AB}$ are given in Appendix \ref{sec:nekrasov-function}. In
this case the norms of the Jack polynomials are given by the vector
parts of the Nekrasov functions:
\begin{equation}
\label{eq:34}
\langle J^{*}_{\vec{A}} , J_{\vec{B}} \rangle = \beta^{-4|\vec{A}|} z_{\mathrm{vect}}(\vec{A},\vec{a}) \delta_{\vec{A},\vec{B}}\,,
\end{equation}
\item \textbf{The ``inversion'' relation.} This relation can be derived by taking the adjoint of the differential
operator $\hat{D}_2$:
\begin{equation}
  \label{eq:28}
  J_{\vec{A}}\left( \left. -\frac{p_k^{(a)}}{\beta} \right| \beta ,\vec{a} \right) =
  (-\beta)^{(N-4) |\vec{A}|} J_{\vec{A}^T} \left(
  p_k^{(a)} \left| \frac{1}{\beta}, -\frac{\vec{a}}{\beta} \right. \right).
\end{equation}

\item \textbf{Cauchy completeness identity.}
  \begin{equation}
  \label{eq:29}
  \sum_{\vec{A}}\beta^{4|\vec{A}|} \frac{J^{*}_{\vec{A}}(p_k^{(a)})
    J_{\vec{A}}(q_k^{(a)}) }{z_{\mathrm{vect}}(\vec{A},\vec{a})} = \exp \left( \beta \sum_{k \geq 1}
   \sum_{a=1}^N \frac{p_k^{(a)} q_k^{(a)}}{k} \right).
\end{equation}
\end{enumerate}

For $\beta \to 1$ the operator $\hat{D}$ becomes the cut-and-join
operator~\cite{cut-and-join} and the generalized Jack polynomials
factorize into products of the Schur polynomials\footnote{Note also
  that for $\beta \neq 1$ the generalized Jack polynomials \emph{do
    not} factorize into the Jack polynomials (as suggested
  in~\cite{Zhang:2011au}), but are instead linear combinations
  thereof.}  $J_{\vec{\lambda}}(\{ p_k^{(a)} \} |\beta \to 1, \{ a_a
\} ) = \prod_{a=1}^N s_{\lambda_{(a)}}(\{ p_k^{(a)}\})$.

We were able to find and check the properties of the generalized Jack
polynomials for $N=3$ up to level $3$. We list here the polynomials at
the level $1$:
{\small
\begin{align*}
  J_{\{\{1\},\{\},\{\}\}} =& \frac{(a_1 - a_2)(a_3 - a_1)
    p_1^{(1)}}{\beta }+\frac{(\beta -1) \left( a_1- a_3\right)
    p_1^{(2)}}{\beta }+\frac{(1-\beta ) \left( \beta -1 -
      a_1+ a_2\right) p_1^{(3)}}{\beta }\\
  J_{\{\{\},\{1\},\{\}\}} =& \frac{\left(a_3-a_2\right) \left(\beta
      +a_1-a_2-1\right) p_1^{(2)}}{\beta }+\frac{(\beta -1)
    \left(\beta +a_1-a_2-1\right) p_1^{(3)}}{\beta }\\
  J_{\{\{\},\{\},\{1\}\}} =& -\frac{\left(\beta
      +a_1-a_3-1\right) \left(\beta +a_2-a_3-1\right) p_1^{(3)}}{\beta }\\
  J^{*}_{\{\{1\},\{\},\{\}\}} =& \frac{\left(\beta -a_1+a_2-1\right)
    \left(\beta -a_1+a_3-1\right) p_1^{(1)}}{\beta }\\
  J^{*}_{\{\{\},\{1\},\{\}\}} =& -\frac{(\beta -1) \left(\beta
      -a_2+a_3-1\right)  p_1^{(1)}}{\beta }-\frac{\left(a_1-a_2\right)
    \left(\beta -a_2+a_3-1\right) p_1^{(2)}}{\beta }\\
  J^{*}_{\{\{\},\{\},\{1\}\}} =& \frac{(\beta -1)
    \left(\beta +a_2-a_3-1\right) p_1^{(1)}}{\beta }+\frac{(\beta -1)
    \left(a_1-a_3\right) p_1^{(2)}}{\beta }+\frac{\left(a_1-a_3\right)
    \left(a_2-a_3\right) p_1^{(3)}}{\beta }
\end{align*}
}
\section{Factorisation of DF integrals}
\label{sec:norm-compl}
The free field representation of the four point conformal block in
Toda field theory is given by the Dotsenko-Fateev integrals:
\begin{multline}
  \label{eq:9}
  \mathcal{B}  = (1 - q)^{(\vec{\alpha}_2 \vec{\alpha}_3)} \Biggl\langle \Biggl\langle \prod_{a=1}^{N-1} \left(
    \prod_{i=1}^{n_{+}^a} \left( 1 - q x_i^{(a)}\right)^{v_{-}^a}
    \prod_{j=1}^{n_{-}^a} \left( 1 - q y_j^{(a)}\right)^{v_{+}^a}
    \prod_{i=1}^{n_{+}^a} \prod_{j=1}^{n_{-}^a} \left(1 - q x_i^{(a)}
    y_j^{(a)}\right)^{2\beta} \right) \times \\
  \times \prod_{a=1}^{N-2} \prod_{i=1}^{n_{+}^a}
  \prod_{j=1}^{n_{-}^{a+1}} \left(1 - q x_i^{(a)} y_j^{(a+1)}\right)^{-\beta}
  \prod_{a=1}^{N-2} \prod_{i=1}^{n_{+}^{a+1}} \prod_{j=1}^{n_{-}^{a}}
  \left(1 - q x_i^{(a+1)} y_j^{(a)}\right)^{-\beta} \Biggr\rangle_{+}
  \Biggr\rangle_{-}
\end{multline}
where the intermediate momentum is determined by the screening
charges: $\vec{\alpha} = \vec{\alpha}_1 + \vec{\alpha}_2 + 2b
\sum_{a=1}^N n^a_{+} \vec{e}_a$. The central charge is $c = (N-1)(1 -
N(N+1)Q^2)$ and $u_{\pm}^a$, $v_{\pm}^a$ are given in
Appendix~\ref{sec:nekrasov-function}. The $\mathfrak{sl}_N$ Selberg
averages are defined as follows:
  \begin{multline}
  \label{eq:25}
  \langle f(x) \rangle_{\pm} = \frac{1}{S} \int\limits_{\mathcal{C}}
  \prod_{a=1}^{N-1} \left[
    d^{n_{\pm}^a}x^{(a)}\prod_{i=1}^{n_{\pm}^a}\left(x_i^{(a)}\right)^{u_{\pm}^a}
    \left( x_i^{(a)} - 1 \right)^{v_{\pm}^a} \prod_{1 \leq i < j \leq
      n_{\pm}^a} \left(x_i^{(a)}
      - x_j^{(a)}\right)^{2\beta} \right] \times\\
  \times \prod_{a=1}^{N-2} \prod_{i=1}^{n_{\pm}^a}
  \prod_{j=1}^{n_{\pm}^{a+1}} \left(x_i^{(a)} - x_j^{(a+1)}\right)^{-\beta}
  f(x)\,,
\end{multline}
and $S$ is the integral without insertions.  The AGT relation requires
the dimensions of the primary fields at $z=1$ and $z=q$ to be
maximally degenerate~\cite{SU(3)}. This implies that $v_{+}^a = v_{+}
\delta^a_1$, $v_{-}^a = v_{-} \delta^a_{N-1}$ in the Selberg
average. For these parameters the expression under the correlator can
be nicely written as an exponential:
\begin{multline}
  \label{eq:18}
  \mathcal{B} = (1 - q)^{(\vec{\alpha}_2 \vec{\alpha}_3)} \Biggl\langle \Biggl\langle \exp \Biggl[ \beta \sum_{k > 0}
  \frac{q^k}{k} \Biggl( \sum_{a=1}^{N-2} \left( p_k^{(a+1)} -
    p_k^{(a)} \right)
  \left(q_k^{(a)} - q_k^{(a+1)} \right) + \\
  +p_k^{(N-1)} \left( -q_k^{(N-1)} - \frac{v_{-}}{\beta} \right) +
  \left( -p_k^{(1)} - \frac{v_{+}}{\beta} \right) q_k^{(1)} \Biggr)
  \Biggr] \Biggr\rangle_{+} \Biggr\rangle_{-} ,
\end{multline}
where $p_k^{(a)} = \sum_{i=1}^{n_{+}^a} \left(x_i^{(a)}\right)^k$,
$q_k^{(a)} = \sum_{j=1}^{n_{-}^a} \left(y_j^{(a)}\right)^{k}$.

Using the Cauchy completeness identity~(\ref{eq:29}) one obtains the
sum of factorised ``plus'' and ``minus'' correlators:
\begin{equation}
  Z = \sum_{\vec{A}}
  \frac{q^{|\vec{A}|} \beta^{4|\vec{A}|}}{z_{\mathrm{vect}}(\vec{A},a)} \left\langle
    J_{\vec{A}} \left(
      -p_k^{(1)} - \frac{v_{+}}{\beta}, p_k^{(2)}-p_k^{(1)},\ldots ,
      p_k^{(N-1)} \right) \right\rangle_{\hspace{-3pt}+} \!\! \left\langle
    J^{*}_{\vec{A}} \left(q_k^{(1)}, q_k^{(1)} - q_k^{(2)},\ldots ,
      -q_k^{(N-1)} - \frac{v_{-}}{\beta}\right)
  \right\rangle_{\hspace{-3pt}-}\!\!. \notag
\end{equation}

To check the AGT conjecture one should ensure that the Selberg
averages of the generalised Jack polynomials reproduce the individual
terms in the Nekrasov function~(\ref{eq:5}):
\begin{gather}
  \label{eq:17}
 \left\langle
    J_{\vec{A}} \left(
      -p_k^{(1)} - \frac{v_{+}}{\beta}, p_k^{(2)}-p_k^{(1)},\ldots , p_k^{(N-1)} \right) \right\rangle_{+} \stackrel{?}{=} \prod_{i=1}^{N}\prod_{f=1}^{N} f_{A_i} (m_f + a_i)\,, \\
   \left\langle J^{*}_{\vec{A}} \left(q_k^{(1)}, q_k^{(1)} - q_k^{(2)},\ldots ,
      -q_k^{(N-1)} - \frac{v_{-}}{\beta}\right) \right\rangle_{-} \stackrel{?}{=} \prod_{i=1}^{N}\prod_{f=N}^{2N} f_{A_i} (m_f + a_i)\,.
\end{gather}
To compute the Selberg averages for several lowest diagrams one should
employ the Virasoro and $W$ constraints~\cite{Itoyama:2011sf}. Using
this method we have checked the relations above for $N=3$ and diagrams
up to order two. We derive the necessary constraints in
Appendix~\ref{sec:w-viras-constr}.

\section{Conclusions and outlook.}
\label{sec:conclusion}

In this Letter we have explicitly found the generalized Jack
polynomials for the group $SU(3)$ and checked the AGT relations for
the $\mathfrak{sl}_3$ Selberg averages on several first levels. We
have also derived the $W$ constraints for the $\beta$-deformed quiver
matrix model, which provide recurrence relations for the correlators
and in principle allow the computation at the arbitrary level.

It would be interesting to investigate the connection between $W$
constraints for the $\beta$-ensembles and the family of the Jack
commuting differential operators (of which $\hat{D}$ is only one
example). It also seems plausible that the general form of these
operators can be found along the lines of~\cite{Nazarov}.

\textbf{Acknowledgements.}  We would like to thank Al. Morozov and A.
Mironov for helpful discussions. A.M. and Y.Z. acknowledge the
hospitality of the International Institute of Physics and DFTE-UFRN in
Natal, Brazil where part of this work was done. Our work is partly
supported by Ministry of Education and Science of the Russian
Federation under contract 8410 (S.~M. and A.~M.), by RFBR grants
13-02-00478, 12-01-33071 mol\_a\_ved (Y.~Z.), 13-02-91371-ST,
12-02-92108-Yaf\_a, 11-01-00962 (A.~M.), 12-01-00525 (S.~M.) and the
Dynasty Foundation (S.~M. and A.~M.).
\appendix

\section{Nekrasov functions and AGT relations}
\label{sec:nekrasov-function}
The Nekrasov partition function for the $SU(N)$ theory with $N_f = 2N$
fundamental hypermultiplets is given by
\begin{equation}
  \label{eq:5}
  Z_{\mathrm{Nek}} = \sum_{\vec{A}} q^{|\vec{A}|}
  \frac{\prod_{i=1}^{N}\prod_{f=1}^{2N} f_{A_i} (m_f + a_i)
  }{z_{\mathrm{vect}}(\vec{A},\vec{a}) }\,,
\end{equation}
where $f_A (x) = \prod_{(i,j) \in A} (x + \epsilon_1 (i-1) + \epsilon_2
(j-1))$, $z_{\mathrm{vect}}(\vec{A},\vec{a}) = \prod_{i,j=1}^{N}
g_{A_i A_j}(a_i - a_j)$ and
\begin{equation}
  \label{eq:2}
  g_{AB} (x)= \prod_{s \in A} \left( x + \epsilon_1 \mathrm{Arm}_A (s)
    -\epsilon_2 \mathrm{Leg}_B (s) + \epsilon_1 \right) \left( x +
    \epsilon_1 \mathrm{Arm}_A (s) -\epsilon_2 \mathrm{Leg}_B (s) -
    \epsilon_2 \right).\notag
\end{equation}

The AGT relations for $N=3$ are:
\begin{gather}
  b(\vec{\alpha}_1 \cdot \vec{e}_a) = u_{+}^a = m_a - m_{a+1} - 1 +
  \beta\,, \qquad \qquad b(\vec{\alpha}_3 \cdot \vec{e}_a) = v_{-}^a =
  -\delta^{N-1}_a \sum\nolimits_{b=1}^{N}
  \widetilde{m}_b\,, \notag\\
  b(\vec{\alpha}_4 \cdot \vec{e}_a) = u_{-}^a = \widetilde{m}_a -
  \widetilde{m}_{a+1} - 1 + \beta \,, \qquad \qquad b(\vec{\alpha}_2
  \cdot \vec{e}_a) = v_{+}^a = -\delta^1_a \sum\nolimits_{b=1}^N
  m_b\,, \notag\\
  \beta n^{1}_{+} = a_2 +a_3 + m_2 + m_3\,, \qquad \qquad \beta
  n^{2}_{+} = a_3 + m_3\,, \notag
\end{gather}
where $\widetilde{m}_a = m_{N+a}$ and $a = 1,2$. Masses $m_a$, the vev
$a_i$ and $\epsilon_{1,2}$ have all the dimension of mass. In this paper
we set the overall mass scale so that $\epsilon_1 = -b^2$, $\epsilon_2
= 1$. In the calculations involving the Jack polynomials we also use the
parameter $\beta = b^2$.

\section{$W_3$ and Virasoro constraints for $\mathfrak{sl}_3$ Selberg
  averages.}
\label{sec:w-viras-constr}
The Virasoro constraints for the Selberg integral are written as follows
{\small
  \begin{gather}
  \label{eq:45}
  \left\langle \left[ \hat{A}(z) - \hat{B}(z) - \beta \rho_1 (z)
      \rho_2 (z) \right] \Phi(w,\bar{w}) \right\rangle = 0 \,, \qquad \text{where}\\
    \hat{A}(z) \Phi(w,\bar{w}) = \Biggl[ (\beta - 1) \partial_z \rho_1
    (z) + \beta \rho_1^2 (z) + \frac{u_1 \rho_1 (z)}{z} + \frac{v_1
      \rho_1 (z)}{z - 1} - \frac{u_1 \rho_1 (0)}{z} - \frac{v_1 \rho_1
      (1)}{z-1}+ \notag\\
    + \sum_{l=1}^{m} \frac{1}{\rho_1 (w_l)} \partial_{w_l} \left(
      \frac{\rho_1 (z) - \rho_1(w_l)}{z - w_l} \right) \Biggr]
    \Phi(w,\bar{w}) \,,\notag\\
  \hat{B}(z) \Phi(w,\bar{w}) = \Biggl[ (1 - \beta) \partial_z \rho_2
  (z) - \beta \rho_2^2 (z) - \frac{u_2 \rho_2 (z)}{z} + \frac{u_2 \rho_2 (0)}{z}
  -\sum_{\bar{l}=1}^{\bar{m}} \frac{1}{\rho_2(\bar{w}_{\bar{l}})} \partial_{\bar{w}_{\bar{l}}}
  \left( \frac{\rho_2 (z) -
      \rho_2(\bar{w}_{\bar{l}})}{z -
      \bar{w}_{\bar{l}}} \right) \Biggr] \Phi(w,\bar{w})\,.\notag
\end{gather}
}
and $\rho_a(z) = \sum_{i=1}^{N_a} \frac{1}{z - x_i^{(a)}}$ and
$\Phi(w,\bar{w}) = \prod_{k=1}^m \rho_1 (w_k)
\prod_{\bar{k}=1}^{\bar{m}} \rho_2 (\bar{w}_{\bar{k}}) $. The $W$
constraints are given by
{\small
  \begin{multline}
  \label{eq:46}
  \Biggl\langle \Biggl[ \beta (\rho_1 (z) \rho_2(z) (\rho_1 (z) -
  \rho_2(z)) + (\beta - 1) \rho_2(z) \partial_z \rho_1(z) +
  \frac{\beta - 1}{\beta} \partial_z \hat{B}(z) + \left( \frac{u_1 +
      u_2}{\beta z} + \frac{v_1}{\beta (z - 1)} \right)
  \hat{B}(z) +\\
  + \left( \frac{u_1}{z} + \frac{v_1}{z - 1}\right) \rho_1(z)
  \rho_2(z)
  + \frac{1}{\beta} \sum_{l=1}^{m} \frac{1}{\rho_1
    (w_l)} \partial_{w_l} \left( \frac{\hat{B} (z)+ \beta \rho_1 (z)
      \rho_2 (z) - \hat{B}(w_l)- \beta \rho_1 (w_l) \rho_2 (w_l) }{z -
      w_l} \right) +\\
  + \frac{1}{\beta} \sum_{\bar{l}=1}^{\bar{m}}
  \frac{1}{\rho_2(\bar{w}_{\bar{l}})} \partial_{\bar{w}_{\bar{l}}}
  \left( \frac{\hat{B} (z) - \hat{B} (\bar{w}_{\bar{l}})}{z -
      \bar{w}_{\bar{l}}} \right) - \frac{u_1}{\beta z} \hat{A}(0) - \frac{u_2}{\beta z}
  \hat{B}(0) - \frac{v_1}{\beta (z - 1)} \hat{A}(1) \Biggr] \Phi(w,\bar{w}) \Biggr\rangle
  =0\,.
\end{multline}
}
If one considers the expansion of the Virasoro and $W$ constraints
$z^{-1}$, $w^{-1}$, $\bar{w}^{-1}$ one obtains the recurrence
relations for the correlators of $p_k^{(a)}$. The coefficients in
front of $z^{-1}$, $z^{-2}$, $z^{-3}$ fix the correlators of
$\rho_{1,2}(0)$, $\rho_1(1)$, $\hat{A}(0)$, $\hat{B}(0)$ and
$\hat{A}(1)$.

{\small
}
\end{document}